# On The Origin Of Unidentified EGRET Gamma-Ray Sources

Olaf Reimer

*Ruhr-Universität Bochum, Theoretische Physik IV, 44797 Bochum, Germany*

**Abstract.** The identification of celestial gamma-ray sources with astronomical objects or object classes has remained the initial and most fundamental key for understanding their physical nature. The observational characteristic of a gamma-ray emitter and the conditions under which an astronomical object is able to produce energetic gamma-ray emission defines the range of candidates available for source identifications. The main obstacle must be seen in the fact that a gamma-ray source location is often imprecise, a flux history could only be established on the basis of weeks, and uncertainties in the gamma-ray observables are considerably large. Therefore coordinated multifrequency follow-up campaigns or spatial-statistical methods are required to assign proper counterpart identifications. Although Active Galactic Nuclei and pulsars are uniquely identified EGRET sources, many other gamma-ray sources still remain unidentified. I will review properties of the population and highlight the characteristics of potential counterparts of the still unidentified gamma-ray sources detected by EGRET.

## INTRODUCTION

With the identification of the Crab pulsar [1] as the first galactic and the quasar 3C273 [2] as the first extragalactic astronomical object emitting gamma radiation, the quest as well as the bounds of source identification at high-energy gamma-ray astronomy became apparent. With disadvantageous spatial localization as result of the photon-limited gamma-ray collecting power in conjunction with a considerable wide point spread function the early detector designs in high-energy gamma-ray astronomy encountered the problematic of unidentified gamma-ray sources instantly. Difficulties in deciding whether an excess is indeed related to an astronomical object or caused by local features in the diffuse gamma-ray emission were apparent in the satellite experiments SAS-2 and COS-B. Only 4 COS-B sources could be identified with unique astronomical objects as counterparts - two pulsars by their timing signature, one AGN due to the remarkable positional coincidence between the radio position and the gamma-ray source in an activity state of this source, and a molecular cloud complex. For the remaining 21 sources no counterpart could be convincingly associated [3] and the concept of a class of unidentified sources was introduced. Later several of these gamma-ray excesses were explained as concentrations of interstellar gas irradiated by cosmic rays [4], indicating that low significance gamma-ray source detections at MeV to GeV energies in our Galaxy are only as good as the understanding of the diffuse emission and, more practically, the quality of the diffuse emission model used in the analysis.

In the 1980s several detection claims of VHE gamma-ray sources were announced from ground based observations, primarily X-ray binaries and pulsars, but also extragalactic objects [5]. Although most of these claims lack verification or appear problematic, no unidentified source has been reported from ground-based gamma-ray telescopes until the turn of the century. This is primarily due to limitations in ground-based gamma-ray astronomy, in particular their pointing strategy to observe candidate objects of VHE gamma-ray emission within a small field of view only. On the other hand, the good spatial resolution and higher gamma-ray rates as possible in observations with imaging atmospheric Cherenkov telescopes, identifications of TeV sources were raised to a new level of significance, immediately resulting in high-significance detections from Mrk 421, and Mrk 501, the confirmation of the Crab Nebula – but also in non-confirmation of various sources reported earlier as detections.

The situation developed dramatically with the launch of the Compton Gamma-Ray Observatory (CGRO) in 1991. All instruments (BATSE [Burst And Transient Source experiment], OSSE [Oriented Scintillator Spectrometer Experiment], COMPTEL [Compton Telescope], and EGRET [Energetic Gamma-Ray Experiment Telescope]) discovered new gamma-ray sources in their respective energy regimes. Despite the many source identifications achieved in the CGRO era, the majority of the high-energy gamma-ray sources remain still unidentified.

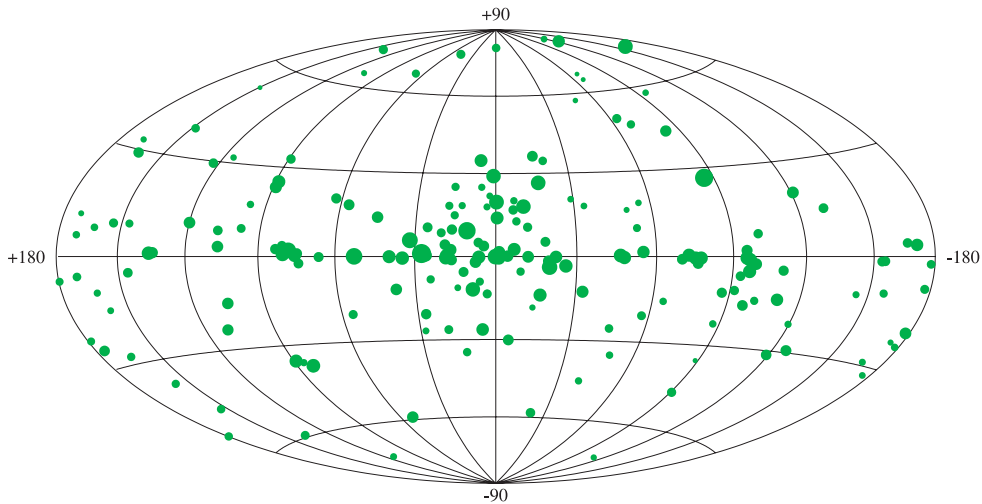

**FIGURE 1.** Unidentified gamma-ray sources from the 3$^{rd}$ catalog of EGRET detected gamma-ray point sources [6].

This fact, equally true for bursts or transients detected by BATSE or gamma-ray point sources discovered by EGRET, is perhaps the biggest challenge in contemporary gamma-ray astronomy: none of these sources have been conclusively identified with an object at other wavelengths, although various objects or object classes have been studied in order to investigate whether they are likely sources of the observed high-energy gamma-ray emission or not. The interest in identifying these sources is based on the following: first and probably most disturbing is the sheer number of unidentified sources – as many as half of the celestial high-energy gamma-ray emitters are still of unknown origin. Second, there is a given chance to discover something new or unexpected from these sources. Even if these sources belong to classes of already known astronomical objects, their identification could significantly improve our understanding of their collective properties if uniquely identified.

With the discovery of gamma-ray emitters like HEGRA J2032+4130 [7] and HESS J1303-631 [8], the problematic of unidentified gamma-ray sources has finally arrived in atmospheric Cherenkov telescope observations. Even in spite of their, in contrast to satellite based gamma-ray astronomy, advantageous precision in source localization and ability to study source morphology, apparently we're confronted with well observed objects not yielding a unique identification with any astronomical object too easily.

## CHARACTERIZATION OF UNIDENTIFIED EGRET SOURCES

Discussing the characteristics of unidentified sources means basically to accumulate all available information about their properties, such as their spatial location, flux history, spectral features, timing information, similarities or differences to known source populations, and suggestive counterparts at other wavelengths. In order to understand their collective properties, a few remarks with respect to observational aspects and instrumental effects of the EGRET telescope and consequences for the results needs to be given. Due to the limited field-of-view of the EGRET telescope all observations are basically pointings towards selected regions of the sky. By co-adding these pointings all-sky surveys were derived as the basis for compilations of point source catalogs. As a consequence, regions of the sky are observed with non-uniform exposure, directly affecting the sensitivity for source detections. Furthermore, the diffuse gamma-ray background is highly structured, strongest along the Galactic Plane, thereby reducing the source detectability towards low galactic latitudes. The detectability of an isolated gamma-ray point source depends on the exposure and diffuse emission in the region of this gamma-ray source, as well as on its flux. This leads to extremely uneven source detectability in EGRET observations. The significance s for detecting an isolated point source [9] can be expressed as

$$s \propto flux \sqrt{\frac{instrumental\ exposure}{diffuse\ emission}} \qquad (1)$$

A visualization of the EGRET detectability matching the $3^{rd}$ catalog is given in Fig.2. Clearly an increase of the detection probability towards higher galactic latitudes is expected; it is seen especially towards positive latitudes. This is due to the particular exposure enhancement from the extensive observation campaigns towards Mrk 421 and 3C279. The lowest chance to detect equally bright sources on the sky by EGRET is directly in the Galactic Plane, due to the pronounced diffuse galactic gamma-ray emission. Not even the many observations towards the Galactic Center could countermeasure this effect in the central part of the Galactic bulge.

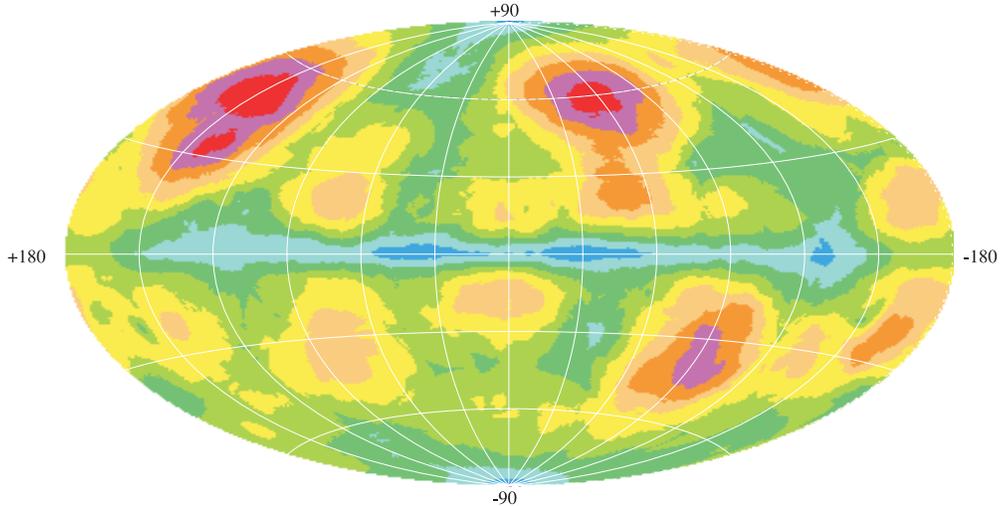

**FIGURE 2.** Sky map of equal source detectability for EGRET observations (E > 100 MeV) matching the 3$^{rd}$ catalog of gamma-ray point sources. Lighter areas indicate reduced source detectability, whereas darker regions exhibit better chances to detect a gamma-ray source at a given intensity. When this figure is compared to Fig.1, it is obvious that the population of unidentified gamma-ray sources in the Galactic Plane is detected against the odds of the detectability by reason of their extraordinary luminosity. This also accounts for the fact that EGRET was increasingly sensitive to less luminous sources away from the Galactic Plane, an indication also used to support a Gould Belt origin of unidentified gamma-ray sources.

Some more peculiarities in the EGRET point source catalogs have to be considered in discussions of source population studies. EGRET point-source catalogs are compiled using two different detection significance thresholds, determined by means of a likelihood method: $\geq 5\,\sigma$ for $|b| < 10°$, and $\geq 4\,\sigma$ for $|b| > 10°$. This results in a non-uniform sensitivity in gamma-ray source detections by EGRET. In addition, the criterion for including a source in the EGRET catalogs is the fulfillment of the detection significance criterion in an individual viewing period, a chosen combination of individual viewing periods, or the total superposition of all viewing periods. Therefore there are cases where a source has been detected if it fulfils the detectability criterion in one single viewing period, although being well beneath such criterion in the analysis of the superposition of all viewing periods. This becomes an essential point in any attempt to compensate the uneven sky coverage of EGRET by means of an exposure correction.

Finally, the number of achieved source identifications depends on the completeness of the catalogs used to check against. Effects such as the zone of avoidance for extragalactic objects at low galactic latitudes caused by interstellar absorption and high background intensity affect radio and optical surveys, therefore reducing the chance to find such counterparts a priori.

With this knowledge one can describe the derived source distributions qualitatively. The identified AGN show an isotropic sky distribution. The excess towards high positive galactic latitudes can be explained by the corresponding detection significance for this region. In contrast, the EGRET-detected pulsars show, although rather statistically limited, a distinct galactic distribution. The distribution of the unidentified sources consists of at least two components. While a distinct excess in the

Galactic Plane indicates a dominant galactic component, a second, apparently isotropic component seems to be superimposed on the galactic distribution. Furthermore, the noticeable enhancement of sources in mid-latitude regions at 3°<|b|<30° has been attributed to the Gould Belt [10, 11, 12]. The Gould Belt, a nearby region of enhanced starbirth activity, is uniquely identifiable due to its asymmetrically inclination of ~20° to the Galactic Plane. Although the indication of a Gould Belt origin of a fraction of the unidentified EGRET sources has been proposed, we currently fail to identify individual sources from this region.

At energies above 1 GeV, the characteristics of gamma-ray source distributions is even more distinguishable. Whereas the AGN population again appears isotropically distributed, the latitude distribution of the unidentified sources shows its excess only in the Galactic Plane. High-latitude unidentified sources above 1 GeV are extremely rare. Among these the absence of bright sources compared to low galactic latitude unidentified sources is especially striking, but reflects the fact that we identify AGN in brightness from top to bottom. Another aspect of high-energy gamma-ray sources is a characterization in terms of their flux variability. Variability is typically addressed on time scales of EGRET viewing periods, typically ranging from 2 to 3 weeks, but investigations for short term variability have been carried out when possible. Variability on any time scales measurable is a typical feature of AGN at gamma-rays. The recent systematic studies of flux variability from EGRET sources [13, 14] reveal and quantify the fact that we identify AGN from their variability signatures: most AGN are clearly variable and thus identifiable as such, whereas the EGRET-detected pulsars are indeed non-variable sources and conclusively identifiable only through periodicity. More strictly, AGN with high average gamma-ray fluxes appear without exception to be variable. High-energy gamma-ray sources can thus be discussed in two categories: variable and non-variable sources. Both are concentrated towards low galactic latitudes. It has turned out that the low-latitude variable gamma-ray sources cannot be easily related to any established gamma-ray source population yet [see Table 1]. In contrast, blazars are known to be highly variable, but could be occasionally found in a quiescent state, distinctly preventing their identification. Short-time variability analysis techniques on time scales of days rather than weeks improve this situation [15].

Further indication for the existence of more than two classes of gamma-ray sources is given by the observations of GROJ1838-04 [16], and 2CG135+01 [17]. The transient character of GROJ1838-04 and difficulty to identify 2CG135+01/3EG J0241+6103 since COS-B lead to the conclusion that objects other than AGN and pulsars might account for these particular sources, and perhaps other unidentified sources too. Table 1 attempts to compare the unidentified sources not only by means of their latitudinal arrangement, but also by addressing their flux variability. Hence, this scheme is suitable for the majority of the unidentified sources, but is expected to be imperfect in some individual cases.

**TABLE 1.** Unidentified EGRET sources and their possible origin

| Variability | Spatial location | |
|---|---|---|
| | Low galactic latitudes | High galactic latitudes |
| Nonvariable | *Galactic objects* *(more distant, higher luminosity)* Pulsars (currently not identifiable as such) SNRs, OB-star associations, Binary systems (orbital modulation not seen yet) Molecular clouds ? | *Galactic objects* *(nearby, lower luminosity)* Runaway pulsars (also radio quiet ?) Gould Belt origin ? Molecular clouds ? *Extragalactic objects* AGN in unfavorable activity states Galaxy Clusters ? Radio galaxies ? |
| Variable | *Galactic objects* Compact objects (microquasars, isolated black holes) ? Binary systems/colliding stellar wind systems (WR-binaries) ? Pulsar Wind Nebulae ? *Extragalactic objects* AGN shining though Galactic Plane | *Galactic objects* Halo origin ? *Extragalactic objects* AGN (currently not identifiable as such) |

## CANDIDATE CAROUSEL OF UNIDENTIFIED SOURCES

In the following the variety of suggested candidates for unidentified EGRET sources are reviewed, starting from the known classes of high-energy gamma-ray emitters to plausible but more speculative classes of astronomical objects not established as gamma-ray emitters to date.

### More Blazar-class Active Galactic Nuclei?

With blazar-class AGN representing the dominant class of identified astronomical objects in the gamma-rays, the most obvious starting point of identifying more gamma-ray sources is to investigate the possibility of their potential AGN origin. Clearly, with the spread in the gamma-ray observables flux variability and spectral slope AGN are ahead in the range of further candidates for still unidentified gamma-ray sources found at all galactic latitudes. With the spatial-statistical assessment of the AGN identifications listed in the 3$^{rd}$ EGRET catalog, Mattox, Hartman & Reimer [18] provided a tool for quantifying BL Lac and flat spectrum radio quasar (FSRQ) associations on the basis of cataloged radio sources (Green Bank 4.85 GHz and 1.4 GHz, PMN 4.85 GHz), making use of the fact that EGRET detected blazars with peak flux above $10^{-6}$ photons cm$^{-2}$ s$^{-1}$ were generally bright (> 1 Jy) radio sources as well. Certainly, with the limitations in completeness and depth in the considered radio catalogs, such methods predictably lack identification power at the low end in radio flux. Sowards-Emmerd et al. [19, 20] and Halpern et al. [21] used additionally optical surveys and performed individual follow up spectral identifications, suggesting a considerable large number of additional blazar identifications among the unidentified gamma-ray sources.

Extensive observation campaigns towards individual objects also yielded new AGN identifications [22, 23]. That a blazar identification in the Galactic Plane can be achieved has been convincingly documented by Mukherjee et al. [24] and Halpern et al. [25] by identifying 3EGJ2016+3657 with the blazar B2013+370.

Comparing log N – log S distributions of identified AGN with the ones of unidentified sources located at high-latitude and low-latitude (Fig.3 [26]), there is strong indication that many of the still unidentified high-latitude sources exhibit different activity states, pointing towards a blazar origin as well. In contrast, the distribution of unidentified source close to the Galactic Plane, the characteristic discrepancy between peak flux and average flux is not apparent, which suggests that the majority among these sources are apparently not blazar-class AGN, although a tendency should be kept in mind that it is less likely to detect soft-spectrum sources in the Galactic Plane due to the pronounced diffuse emission.

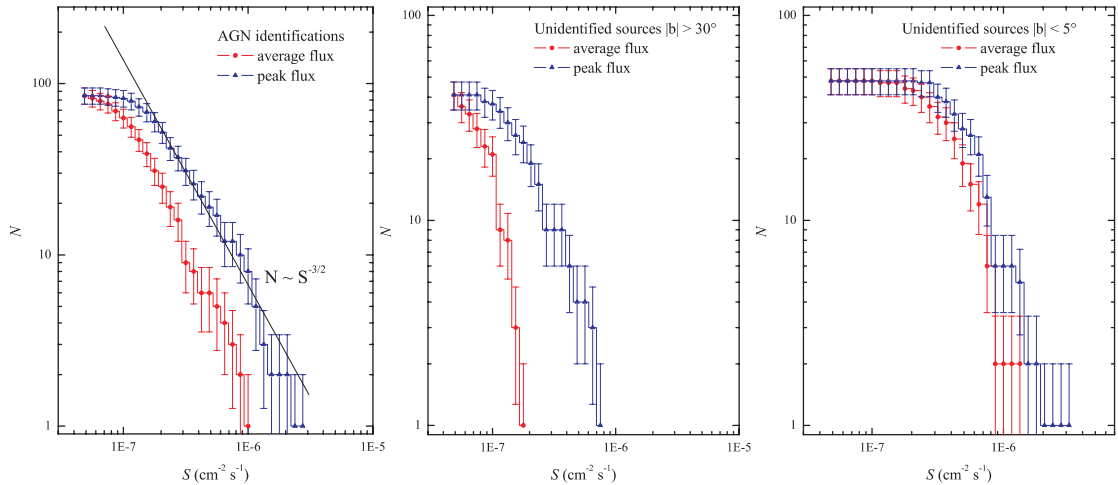

**FIGURE 3.** Log N - log S distributions of identified AGN, unidentified gamma-ray sources at high galactic latitudes (|b| > 30°) and low galactic latitudes (|b| < 5°), shown with 4-year average as well as peak flux. The similarity between peak and average fluxes in the AGN population and the high latitude unidentified sources indicate that a large fraction among these sources very likely awaits their blazar identification. The log N – log S distribution for galactic sources appears distinctly different – rarely gamma-ray sources are detected in extreme variability states. This is at least partly due to the suppressed source detectability in the Galactic Plane, in particular amplified for gamma-ray emitters with softer spectrum than the galactic diffuse emission. The apparent flattening at lower fluxes than a few $10^{-7}$ cm$^{-2}$ s$^{-1}$ is an indication of the instrumental and observational inability to detect fainter sources equally sensitive at different regions in the sky. This effect can be investigated more closely if the inhomogeneous application of source detection criteria in the 3$^{rd}$ EGRET catalog as well as the irregular source detectability (Fig.2) is accounted for accordingly. Unfortunately, the EGRET data barely allow more precise spatial comparisons considering the low source statistics and already apparent systematic biases.

## More Energetic Pulsars Or Pulsar Wind Nebulae?

With the still ongoing expansion of the radio pulsar catalogs from the Parkes Multibeam Pulsar Survey (PMPS) and the Arecibo Drift-Scans the number of pulsars known at the time of the CGRO from the Princeton catalog has been significantly increased, thus new positional coincidences between unidentified gamma-ray sources and radio pulsars are expected – and consequently found [27]. Although contemporaneous timing solutions for the EGRET observations performed a decade ago are not possible, a number of promising positional coincidences between newly discovered pulsars and unidentified EGRET sources await their proof-of-identification in the GLAST era. These candidates are similar in their age and spin-down luminosity with the few established gamma-ray pulsars, so no new physical assumptions are implicated from such identifications. However, there are prospects to distinguish radio-loud and radio-quiet pulsars, currently only possible in population synthesis studies [28, 29, 30]. Significant differences in the expected number of detectable radio-quiet and radio-load pulsars will be not only be a powerful test to distinguish different pulsar emission models, but also predict the range of anticipated pulsar parameters beyond the coverage achieved in gamma-ray observations. The polar-cap and outer-gap models in their variants also predict different numbers. Of interest here is a comparison of the number of pulsars EGRET should have seen: McLaughlin and Cordes estimated ~20 pulsars to be detectable by EGRET [29], Gonthier et al. ~25 [31], with 2-3 times more radio-loud than radio-quiet pulsars. Even more specific Harding & Zhang account for ~11 neutron stars from the Gould Belt alone [32], emphasizing the distinction between on- and off-beam emission. Outer-gap models generally invert the ratio between radio-loud and radio-quiet pulsars, thus predicting considerably larger numbers of pulsars among the unidentified EGRET soures [33, 28]. More recent predictions impose up to 75 pulsars in the Galaxy with ~20 located in the Gould Belt [34].

Progress in identifying pulsar candidates is not exclusively driven from radio observations – also deep multifrequency observations on individual gamma-ray sources yielded conclusive results: The identification of 3EGJ2227+6122 with PSRJ2229+6114 [35], originating from a follow-up on the earlier found association with an X-ray source RX/AXJ2229.0+6114, and the very suggestive identification of 3EG J1835+5918 with an isolated neutron star candidate RXJ1836.2+5925 [36, 37].

When gamma-ray sources in spatial coincidence with energetic pulsars exhibit evidence for gamma-ray flux variability, synchrotron nebulae (PWN) powered from the pulsar are preferred alternatives to pulsar identifications. PWNe have emerged as counterparts at TeV energies as in case of the Crab, Vela and PSRJ1709-4429 nebulae, and a number of associations are indicated. Among them we find several PWN/SNR associations, which leave a certain level of ambiguity in these associations. Nevertheless, associations of PWNe with unidentified EGRET sources have been suggested in cases of 3EGJ1013-5915, 3EGJ1410-6147, 3EGJ1837-0423, 3EGJ1856+0114, 3EGJ2021+3716, and 3EGJ2227+6122 [38]. The plausibility of these associations will be verifiable in the GLAST era.

Finally, detecting radio pulsars in blind fold searches did not yielded new pulsar candidates from the EGRET data anymore [39].

# Supernova Remnants, The Birthplaces Of Pulsars?

The association between gamma-ray sources and supernova remnants (SNRs) has a long history and dates back to the COS-B era [40]. Triggered from the many more source detections by EGRET, SNRs remain statistically significant correlated with unidentified sources at the 5 s level [41]. Although many of these spatial associations have been extensively investigated in multifrequency observations utilizing the most powerful astronomical facilities, we still lack even a single unambiguous identification of a SNR at GeV energies − we are struck with unidentified sources in positional coincidence with supernova remnants. In contrast, with the H.E.S.S. observations of SNR RXJ1713.7-3946 (G347.3-0.5) showing a morphological resolved image of the SNR at TeV energies, there can be no doubt that SNRs are indeed emitters of detectable energetic gamma-radiation [42]. Thus, the lack of conclusive SNR identifications among the EGRET detected gamma-ray sources may be at least partly attributed to limitations in the angular resolution in satellite-based gamma-ray experiments. However, there are also differences in the observed gamma-ray characteristic apparent. In several cases (3EGJ0010+7309/G119.5+10.2 [CTA1], 3EGJ0617+2238/G189.1+3.0 [IC443], 3EGJ1800-2338/G6.4-0.1 [W28], 3EGJ1856+0114/G34.7-0.4 [W44], and 3EGJ2020+4017/G78.2+2.1 [γCygni] the gamma-ray source locations are characterized by remarkable mismatches in either the source location in respect to the morphology of the remnant known from radio or X-ray observations or a discrepancy between the extension of a remnant and a well localized high-energy gamma-ray source inside the remnant, not correlating with any dominant remnant features like shells or rims. By investigating the spectra of such associations (Fig.4 [43]) deviations from a single power law fit are seen, as more pronounced as better the source have been observed.

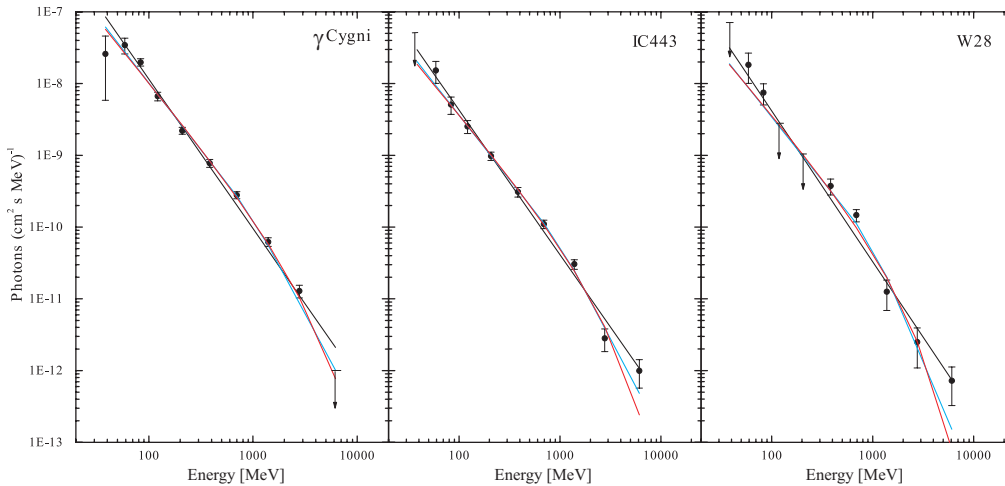

**FIGURE 4.** Photon energy spectrum of three EGRET sources believed to be associated with SNRs: 3EGJ2020+4017, spatially coincident with SNR G78.2+2.1 (γCygni), 3EGJ0617+2238, spatially coincident with SNR G189.1+3.0 (IC443), and 3EGJ1801-2312, spatially coincident with SNR G6.4-0.1 (W28), each shown with three different representations of a spectral fit. In all of these EGRET sources, a hard spectrum as well as the evidence of a spectral cutoff at GeV energies indicates rather a pulsar origin of the observed GeV emission than a manifestation of particle acceleration in the respective SNR.

The fact, that we do not see a morphological similarity in the associations of SNRs at GeV gamma-ray sources in conjunction with the indication of spectral cutoffs at GeV energies is suggestive of another potential identification: young pulsars not having kicked-off too far from their birthplace. In the above mentioned positional coincidences we find X-ray point sources exhibiting the characteristic features of pulsars, although we're currently unable to verify their periodicity fingerprint, either because of the unknown ephemerides and the photon-limited character of satellite-based gamma-ray observations or because of geometrical reasons, where one simply misses the pulsars beam towards us [32, 44]. Clearly, the different spectral emission characteristics of a gamma-ray source of either pulsar or SNR origin can be solved in observation by GLAST-LAT.

## Massive Stars In Regions Of Enhanced Starformation Or Dense Molecular Clouds?

Similar to the situation concerning the SNR associations coincidences between gamma-ray sources and regions of enhanced star formation and containing massive stars have been noticed [40, 45]. On the basis of the 3EG catalog, 26 spatial associations with OB stars were suggested by reason of positional coincidence with a gamma-ray source [46]. When comparing the luminosity distribution obtained by using OB-star association distances, the luminosities are consistent with the longitudinal, latitudinal, and flux distribution of unidentified EGRET sources in the Galaxy. From individual coincidences conclusions for the entire class are predicted by modeling our local neighborhood. However, with their role as possible tracer for pulsars a OB-star association nature of an unidentified gamma-ray source can only be revealed if a SNR, PSR or OB-association origin can be convincingly disentangled, i.e. by variability or lack-of-variability, source extension or morphological similarities in multifrequency observations. Here, in particular the unidentified source HEGRA J2032+4130, located in the vicinity of a well-known region of massive star formation Cyg OB2, may be the prototype for such association.

Another class of objects tested for positional coincidence with unidentified gamma-ray sources are Wolf-Rayet (WR) stars. These objects might be capable of producing gamma-ray emission well into the >100 MeV range from interactions of their stellar winds [47, 48] and non-thermal emission has been indeed seen up to X-ray energies in a few cases. By comparison of coincidences between unidentified EGRET sources and 160 catalogued WR stars these objects are also suggestive as possible counterparts of unidentified gamma-ray sources in our Galaxy. However, the detailed study by Romero et al. [46] comparing WR-stars with unidentified 3EG catalog sources reported that only a small number of individual WR-stars can be actually considered as candidates from positional coincidences. Because several WR-stars are binary systems, particle acceleration in colliding wind scenarios appear to be a viable physical explanation for such counterparts, so they will be mentioned in the next section.

Dense molecular clouds are also suggested as counterparts for individual unidentified EGRET sources [49, 50]. Whereas the fundamental emission mechanism is long since at hand [i.e. 51, 52], individual associations still remain at the level of positional coincidences. Deficiencies in the diffuse gamma-ray emission model used in the EGRET likelihood source finding algorithm leave sufficient room for alternative explanations.

## Binary Systems With Compact Objects (MicroQSOs) Or Colliding Stellar Wind Systems?

With the confirmation of VHE gamma-ray emission for the binary system PSRJ1259-63/SS 2883 [53, 54] the first object from a new class of gamma-ray emitting objects could be established: astronomical objects in close binary systems. At GeV energies, several binary systems are proposed as counterparts for unidentified EGRET sources, with constituents spanning the range from compact objects to massive stars. First, a very suggestive analogy exists between AGN harboring a supermassive black hole and exhibiting large-scale jets, and galactic microquasars [55] believed to consist of compact objects which accrete from massive companion stars and exhibiting parsec-scale jets. Observationally, we still fail to identify any of the highly variable sources located within the Galactic Plane, therefore microquasars are appealing candidate sources. Indeed, currently three positional coincidences (Fig.5) between an EGRET source and a microquasar have been reported [17, 56, 57].

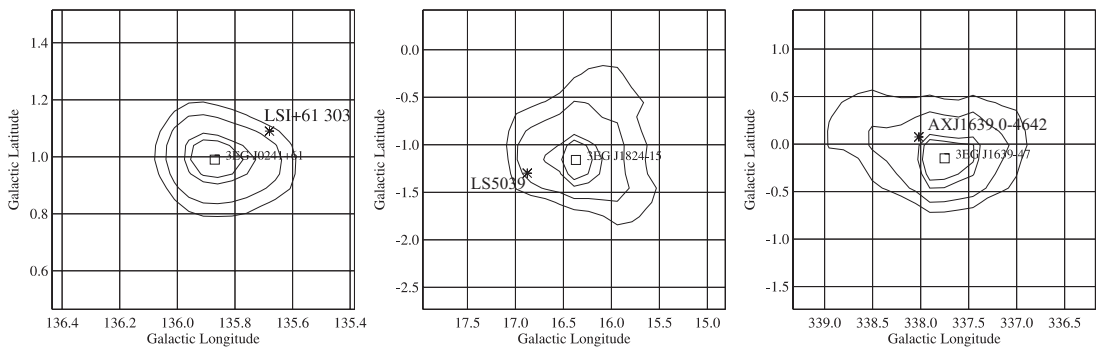

**FIGURE 5.** Source location maps as given in the 3$^{rd}$ EGRET catalog. The respective microquasars have been noticed as potentially associated with high-energy gamma-ray emission. Contours represent the 50**%**, 68**%**, 95**%** and 99**%** statistical probability that a single source lies within the given contour; the cross denotes the position of the microquasar [58].

Whereas 2CG135+01/3EGJ0241+6103 has a rather long observational record already and is suggested to be associated with the eclipsing high-mass binary system LSI+61°303, the other candidates are newly discovered high mass X-ray binary systems (HMXBs). Gamma-rays can arise in Inverse Compton processes when the relativistic jet electrons interact with external photons from the massive companion or from the decay of $\pi^0$'s in interactions between jet protons and hadrons from the wind of the companion star. Microquasars naturally exhibit the gamma-ray emission characteristics not found in any other galactic candidate source population yet: rapid, unpredictable source activity, perhaps also modulated by orbital motion, relativistic particle populations in a jet and a reservoirs for accretion in form of massive companion stars.

Finally, it should be mentioned, that with the HMXB Cyg X-3 a long disputed candidate gamma-ray emitter lies in the vicinity of the EGRET source 3EGJ2033+4118, but neither the 4.8 h nor the 12.59 ms periodicity could have been found [59]. Also, the source 3EGJ0634+0521 has been found positional coincident with SAXJ0635+0533 [60], an X-ray pulsar in a binary system with an 11.2 day orbital periodicity [61].

Less energetically violent but also well in the range of potential counterparts of unidentified gamma-ray sources are binaries in colliding stellar wind systems, with WR140, and WR 147 the best examples of non-thermal activity associated with colliding wind WR+O binaries. Several positional coincidences have been noticed by Romero et al. [46], with 3EGJ2022+4317/WR140, 3EGJ2021+3716/ WR142, and 3EGJ2020+4317/Cyg OB2#5 of particular interest concerning potential high-energy gamma-ray emission [47, 48, 62]. The chance to observe orbital modulations in their gamma-ray intensity will be the decisive proof of such associations.

## Seyferts, Starburst, Or Radio Galaxies?

In the 3$^{rd}$ EGRET catalog, only Cen-A and the Large Magellanic Cloud have been reported as non-blazar class identifications of extragalactic gamma-ray emitters. Other galaxies have been studied in the EGRET data as well, in particular seeking advantage from statistical gain achievable in source stacking techniques. Individually, none of the prominent starburst galaxies have been found in the gamma-ray data [63], and also LIGs and ULIRGs cannot account for counterparts of EGRET sources [64, 65]. In contrast, individual radio galaxies found in the vicinity of unidentified EGRET sources (3EGJ1621+8203/NGC 6251 [66], 3EGJ1735-1500 [67], 3EGJ0416+3650/ 3C111 [68], ) as well as a Seyfert 1 (3EGJ1736-2908/GRS 1734-292 [69]) have been proposed as potential counterparts albeit a comprehensive population study by Cillis et al. [70] did not revealed radio galaxies nor Seyferts as likely candidates for high-energy gamma-ray sources exceeding fluxes of ~ $10^{-8}$ photons cm$^{-2}$ s$^{-1}$ (E > 100 MeV). It should be noted that there needs to be drawn a distinction between a positional coincidence and a counterpart identification. Counterpart identification requires positional agreement with an identified source, a plausible gamma-ray emission mechanism and consistent multifrequency characteristics (both temporal and spectral). Consequently, the proposed associations with radio or Seyfert galaxies must await their verification in the GLAST era.

## Galaxy Clusters?

Recently, clusters of galaxies have been suggested for association with unidentified gamma-ray sources, in cases of individual associations [71], in correlation with the population of unidentified EGRET sources [72], and also as unresolved gamma-ray excesses accounting for the majority of the observed extragalactic gamma-ray emission [73]. Although galaxy clusters are very appropriate candidates for the next-generation gamma-ray instrumentation, in particular massive and nearby clusters or cluster merger systems, the suggested associations leave more questions unanswered than answered: the proposed counterparts are neither the most obvious galaxy clusters predicted to be detectable at high-energy gamma-rays nor it can be understood why several inconspicuous Abell-clusters are the first ones to be seen at gamma-rays when the number of prominent and well-studied galaxy clusters exhibiting evidence for non-thermal radiation processes at radio, EUV or hard X-ray wavelengths is not yet detectable. For a flux limited sample of the X-ray brightest galaxy clusters only upper limits could be determined using the nine years of EGRET observations [74]. The correlation between Abell-clusters and unidentified gamma-ray sources at $|b| > 10°$ remains at the level of statistical insignificance when autocorrelation in the large sample is taken into account [74]. Thus, clusters of galaxies may not account for a sizable quantity among the unidentified EGRET sources.

## CONCLUSION

A vast number of unidentified EGRET gamma-ray sources still resist their conclusive identification although many viable candidates are at hand. A capable instrument like GLAST-LAT will enable unambiguous identifications of all cataloged gamma-ray sources of the EGRET era but likely encounter an even larger population of unidentified GLAST sources at the $\sim 10^{-9}$ photons cm$^{-2}$ s$^{-1}$ level. Imaging atmospheric Cherenkov telescopes provide complementary ways to investigate high-energy gamma-ray emitters with high photon rates and arcmin source locations. Despite the success of multifrequency identification campaigns in achieving conclusive individual source identifications, the better capabilities of the next-generation instrumentation may allow identifications of a large fraction of high-energy gamma-ray sources on its own. We should not worry if the existence of unidentified gamma-ray sources will persist unless we fail to identify intriguing individuals or entire new classes of astronomical objects among them.

## ACKNOWLEDGMENTS

O.R. acknowledges support from the BMBF trough DLR grant QV0002.